\documentclass[%
 reprint,
superscriptaddress,
 amsmath,amssymb,
 aps,
]{revtex4-2}

\usepackage{graphicx}
\usepackage{dcolumn}
\usepackage{bm}
\usepackage{subfigure}
\usepackage{multirow}
\usepackage{amsmath,epsfig,amssymb,subfigure,bm,dsfont}
\usepackage{lipsum}

\begin{document}

\preprint{APS/123-QED}

\title{Deterministic generation of arbitrary n-photon states in an integrated photonic system}

\author{Fan Xing}
\affiliation{ State Key Laboratory of Optoelectronic Materials and Technologies, School of Physics, Sun Yat-sen University, Guangzhou 510275, China}

\author{Zeyang Liao}%
\email{Corresponding author: liaozy7@mail.sysu.edu.cn}
\affiliation{ State Key Laboratory of Optoelectronic Materials and Technologies, School of Physics, Sun Yat-sen University, Guangzhou 510275, China}%

\author{Xue-hua Wang}
\affiliation{ State Key Laboratory of Optoelectronic Materials and Technologies, School of Physics, Sun Yat-sen University, Guangzhou 510275, China}

\date{\today}

\begin{abstract}
Quantum light sources play a vital role in various aspects of quantum information science, but on-demand high-efficient generation of arbitrary multiphoton states which can be easily integrated is still challenging.  Here, we propose a chip-integrable scheme to deterministically generate a group of $n$ photons with very high fidelity based on the long-range collective interaction between the emitters mediated by the waveguide modes. The $n$ photons are shown to be emitted in a bundle while two successive $n$-photon bundles tend to be antibunched which can behave as an $n$-photon gun. Our results here can find important applications in the areas such as photonic-chip-based quantum information processing and quantum metrology.
\end{abstract}
\maketitle

\section{Introduction}

The classical coherent light source like laser has been extensively applied in many areas and played important roles in the modern information science \cite{Maiman1960,song2019}. Similarly, quantum light sources such as single photon sources and entangled photon sources have become key ingredients for photonic quantum technologies \cite{Trivedi2020,kok2007,Lund2014,Ekert1991,Hensen2015,Nilsson2013}. The single photon sources can be heralded generated by the parametric down conversion \cite{Heidmann1987,Harder2016,Han-Sen2020,madsen2022} or four-wave mixing processes \cite{paesaniStefano2020}, but these methods are usually probabilistic and their efficiencies are usually very low. The photon blockade effect can also be exploited for single photon generation in the cavity quantum electrodynamics (QED) system, but the efficiency is also not very high \cite{muller2015,hamsen2017,chen2022,mckeever2004,darquie2005,Trivedi2019,lu2021}. During the past two decades, the solid state quantum emitters like semiconductor quantum dots (QDs) has been shown as very promising material platform for on-demand high efficiency and high quality single photon sources \cite{zhai2022,laferriere2020,Michler2000,buckley2012,Ding2016,aharonovich2016,tomm2021} and entangled photon sources \cite{muller2014,liujin2019}. The single photon sources generated by a high-quality QD have been temporally multiplexed to demonstrate the quantum advantages in Boson sampling problem \cite{Loredo2017,wanghui2017}.

In recent years, multiphoton Fock states or bundle emissions where the system emits a bundle of strongly correlated n photons have attracted much attentions due to their fundamental applications in the areas such as quantum information protocols \cite{haroche2006}, N-photon lasers and photon guns \cite{lounis2005}, quantum lithography and metrology \cite{giovannetti2004,giovannetti2011}, medical applications with high resolution and minimum harm to tissue \cite{denk1990,horton2013}, biological photoreceptors with sensitivities greater than classical light \cite{sim2012,ball2011}. The Fock state can be conditionally generated via the state collapse from a coherent or thermal state \cite{Guerlin2007,Fran2001}, but this process is not deterministic and the probability decreases rapidly when n increases. By accurately controlling the cavity-qubit coupling strength and interaction times, multiple photon states can also be prepared  \cite{Law1996,Brown2003,liuyuxi2004,hofheinz2008,xiakeyu2012}. Mu{\~n}oz et al. showed that a strongly coupled cavity-atom system can form a series of anharmonic dressed states and if the pumping frequency satisfied the right conditions the system can emit n-photon in a bundle \cite{munoz2014}. Actually, if the usual Jaynes-Cummings (JC) Hamiltonian in the cavity-QED system can be generalized to the higher-order n-photon JC Hamiltonian \cite{Sukumar1981,Singh1982,Jiang2023}, the coupled system can oscillate between the ground state and the n-excitation state (i.e., super-Rabi oscillations) and n-photon state can be also generated \cite{VillasBoas2019,yinxianli2022}. Taking advantage of the nonlinear dynamics of the Cooper-pair tunneling or the ultrastrong coupling, the effective Hamiltonian of the circuit-QED system can be described by the n-photon JC model and antibunded N-photon bundles can also be generated \cite{bin2021,bin2020,mashengli2021,menard2022}. However, when n is relatively larege, the efficiencies of the above methods are usually low. Uria et al. showed that a cavity field initially in a coherent state interacting with an atom can evolve to a quantum state close to a Fock-like state at a certain interaction time, and they numerically showed that $n\sim 100$ Fock-like state with $70\%$ fidelity can be generated \cite{uria2020}. However, the generated state in this method is not a Fock state but a displaced Fock state. By coupling the field to a multilevel atom, multiple photons can also be generated \cite{sanchez2016,lu2022,groiseau2021}, but the number of photons is largely restricted by the available atomic levels.  Gonz{\'a}lez-Tudela et al. proposed an interesting method for deterministic generation of arbitrary multiphoton states in a waveguide-QED system by pumping the system into an N-excitation many-body collective states through carefully-designed pulse sequences which is not very friendly for experimental implementation \cite{gonzalez2015,Gonzalez2017}.

In this letter, we propose an alternative scheme to generate arbitrary n-photon state on demand in an integrated photonic system which is relatively easy to be implemented. In our scheme, we employ the long-range dipole-dipole interaction mediated with waveguide modes where the emitters with large separation can still have strong collective interactions \cite{Liao2015,Hood2016,Cheng2017,Douglas2015,Corzo2019,Xing2022}. Since the emitters here need not to be very close, we can selectively pump specific number of emitters into the excited state $|e\rangle$ by applying multiple ultrashort coherent $\pi$ pulses whose duration is much less than the collective decay time of the emitters. Then they can cascade down to the ground state through the superradiant pathway and emit certain number of photons with high correlations into the waveguide. Our results show that the photons are emitted in a bundle and the photons in different bundles are antibunched. The proposed scheme can find important applications in the chip-integrated photonics quantum devices \cite{Zheng2013,Lodahl2015,GU2017,Liao2018,JWang2020, Bogaerts2020, Arrazola2021,Elshaari2020,Pelucchi2022}.  

 This article is organized as follows. In Sec. II, we discuss our scheme and the theoretical calculation methods. In Sec. III, we numerically demonstrate the validity of our scheme through three-atom examples. Finally, we summarize our results.

\section{MODEL AND THEORY}

\begin{figure}[tbp]
\label{Fig1} \centering\includegraphics[width=8.5cm]{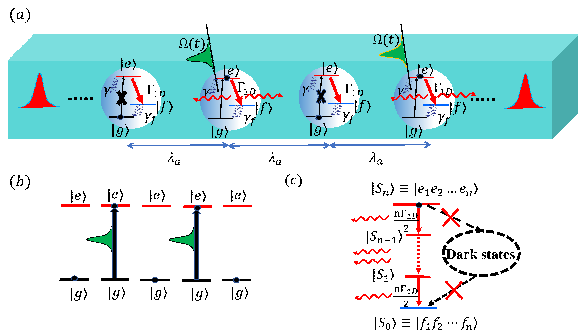}
\caption{(Color online) (a) Schematic diagram for generating n-photon bundle states in 1D waveguide-QED system where N multi-level emitters couple to the waveguide. (b) The emitters are selectively pumped from the ground state $\vert g \rangle$ to the excited $\vert e \rangle$ by external driving $\pi$ pulses. The emitters without being pumped remain in the ground state. (c) The emitters being pumped to the excited state $\vert$e$\rangle$ can couple to the waveguide and they can collectively decay to the intermediate state $\vert$f$\rangle$ and release n highly correlated photons into the waveguide. }
\label{1}
\end{figure}

The schematic model is shown in Fig. 1(a) where $N$ identical multi-level emitters couple to a 1D single-mode waveguide. Among these atomic levels, only the transition between $\vert$e$\rangle$ and $\vert$f$\rangle$ with transition frequency $\omega_{ef}$ can couple to the waveguide modes and other transitions are decoupled from the waveguide but they can couple to the external nonguided fields. The emitters are initially in the ground state $\vert$g$\rangle$ which decouple to the waveguide and can be selectively pumped to the excited state $\vert$e$\rangle$ by an external laser $\pi$ pulse whose frequency is tuned to be resonant with the $\vert$g$\rangle$ $\rightarrow$ $\vert$e$\rangle$ transition (Fig. 1(b)). The emitters without being pumped will stay in the ground state $\vert$g$\rangle$ which are decoupled from the waveguide. An emitter in the excited state $\vert$e$\rangle$ can either decay to the state $\vert$f$\rangle$ with rate $\Gamma_{1D}$ and emit a photon into the waveguide or decay back to the ground state $\vert$g$\rangle$ with rate $\gamma$ and emit a photon to the free space. For a single-emitter case, the collection efficiency is given by $\Gamma_{1D}/(\Gamma_{1D}+\gamma)$. For the n-emitter excitation case, the emitters can collectively decay to the state $\vert$f$\rangle$ and emit n photons into the waveguide. The collection efficiency can be enhanced which is given by $n\Gamma_{1D}/(n\Gamma_{1D}+\gamma)$.  The emitters in the state $\vert$f$\rangle$ can cascade down to the ground state $\vert$g$\rangle$ through certain intermediate states with effective decay rate $\gamma_{f}$. By repeating the above procedures, we can produce a sequence of photon pulses with each pulse containing n-photon bundle. This is the basic principle of our scheme.

To deterministically generate pure $n$-photon states, it is critical to selectively pump $n$ emitters into the excited state $\vert e \rangle$ with high fidelity and the other irrelevant emitters are decoupled from the waveguide. To achieve this, we apply multiple focused laser beams to the target emitters from the direction perpendicular to the waveguide. The emitter distance is chosen to be much larger than the diffraction limit which is about half wavelength to avoid cross excitation. In addition, to ensure that the selected emitters are prepared in the excited states $|e\cdots e\rangle$ with very high fidelity, the pulses are chosen to be $\pi$ pulses and their time duration should be much less than the collective decay time of the emitters $(1/n\Gamma_{1D})$. Under this condition, the emitters almost do not decay during the pumping process and after that they decay without pumping to avoid the occurring of reexcitations. 

The emitters in the excited $\vert$e$\rangle$ can couple to the waveguide and they can form collective states due to atom-atom interaction mediated by the waveguide modes. By tracing out the effects of the waveguide field, the effective Hamiltonian of the emitters in the subspace of $\vert$e$\rangle$ and $\vert$f$\rangle$ is given by \cite{Liao2015,Hood2016,Liao2016}
\begin{eqnarray}
H_{eff}^{(ef)}\text{=}\ \hbar\omega_{ef}\sum_{j}\sigma_{ef}^{j}\sigma_{fe}^{j}-i\frac{\hbar\Gamma_{1D}}{2}\sum_{j,l=1}^{N}e^{ik_{a}z_{jl}}\sigma_{ef}^{j}\sigma_{fe}^{j},
\label{0}
\end{eqnarray}%
whose real part gives the collective effective energy and the imaginary part describes the collective decay rates. $\sigma_{ef}^{j}=\vert e \rangle_{j} \langle f \vert$ is the operator describes the $\vert f \rangle \rightarrow \vert e \rangle$ transition of the $j$th emitter. $z_{jl}=|z_{j}-z_{l}|$ is the distance between the $j$th and $l$th emitters. By diagonalizing $H_{eff}^{(ef)}$, we can obtain the effective collective eigen-energies and eigenstates of the emitter system in the subspace spanned by $\vert e \rangle$ and $\vert f \rangle$. Due to the collective interactions, some states are superradiant and others are subradiant. Here, we consider the case when the emitter distances are integer number times of resonant wavelength under which the subradiant states completely decouple from the waveguide field and only the superradiant states can couple to the waveguide. Without loss of generality, here we assume that the distance between the nearest-neighbor emitters is the resonant wavelength $\lambda_a$ corresponding to the transition frequency $\omega_{ef}$. The superradiant states are given by $\vert S_{m} \rangle=\frac{1}{\sqrt{n}}sym\{\vert e \rangle^{\otimes m}\vert f \rangle^{\otimes(n-m)}\}$ which is the symmetric superposition of the states with $n$ emitters being in the state $\vert e \rangle$ and the other $n-m$ emitters being in the state $\vert f \rangle$. The eigenenergy of the state $\vert S_{m} \rangle$ is $m\omega_{ef}$ with collective decay rate $\Gamma_{m}=m(n-m+1)\Gamma_{1D}/2$ where $m=0,1,...,n$ (Fig. 1(c)). When $m=0$, $\vert S_{0} \rangle=\vert f...f \rangle$ is the lowest energy eigenstate in the subspace while when $m=n$, $\vert S_{n} \rangle=\vert e...e \rangle$  is the highest energy eigenstate with decay rate $n\Gamma_{1D}/2$ which is $n$ times larger than that of the single emitter decay rate. For other values of $m$, the collective decay rates can be larger than $n\Gamma_{1D}/2$. If the emitter system is prepared in the state $\vert S_{n} \rangle$ with high fidelity, it can cascade down to the ground state through the superradiance pathway and deterministically emit $n$ photons with high purity. Due to the superradiant effect, the emitted $n$ photons are bounded in the time and space domain.

\section{NUMERICAL SIMULATION}

\begin{figure}[tbp]
\label{Fig2} \centering\includegraphics[width=9cm]{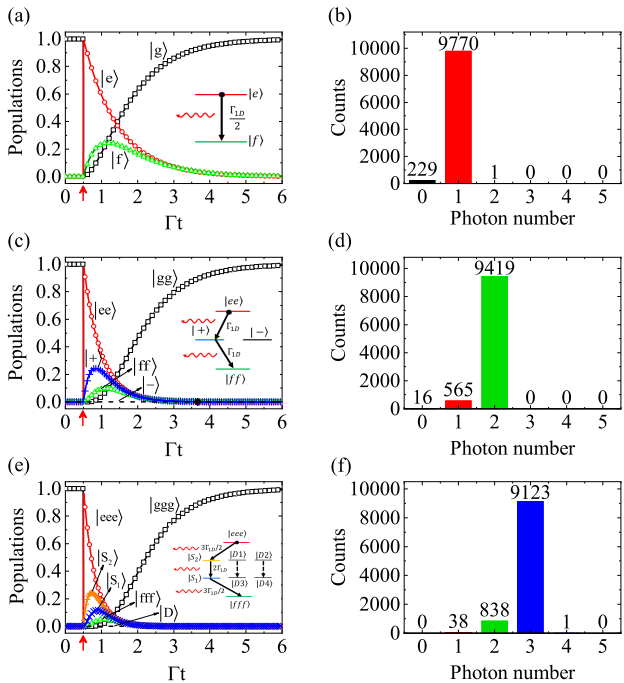}
\caption{(Color online) Population dynamics and photon count statistics for the one-excitation case (a, b), two-excitation case (c, d), and three-excitation case (e, f). }
\label{2}
\end{figure}

To prove the validity of our scheme, we can numerically calculate the population dynamics and the photon statistics either by solving the master equation  \cite{liao2020} or using the quantum Monte Carlo method (see the supplementary material)\cite{zhang2018,plenio1998,molmer1993}. Here as an example we consider the case when three emitters couples with a 1D waveguide and compare the results when one or two or three emitters are excited. The nearest-neighbor separation between the emitters is set to be $\lambda_{a}$. The decay rates are given by $\gamma=0.05\Gamma_{1D}$ and $\gamma_{f}=2\Gamma_{1D}$. Initially, all the emitters are in the ground states. We then apply multiple coherent driving $\pi$ pulses with time duration much less than the collective decay time $1/n\Gamma_{1D}$ ($n=3$ in this example)  to selectively excite certain number of emitters. In the following numerical simulations, we assume that the driving pulse is Gaussian $\pi$ pulses with spectrum width $\Delta=200\Gamma_{1D}$ and average photon number $\bar{n}_{d}=4182$.

\begin{figure}[tbp]
\label{Fig3} \centering\includegraphics[width=9cm]{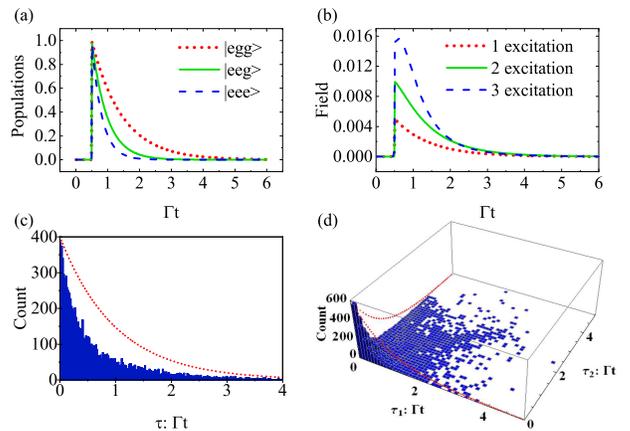}
\caption{(Color online) (a) Comparison of the population decay dynamics and the output photon intensities of the cases when single, double, and three emitters are excited. (c, d) Statistics of the time intervals of the 9419 two-photon events and 9123 three-photon events observed in Fig. 2.}
\label{3}
\end{figure}

The population dynamics of the emitter system as a function of time and the photon statistics for the three excitation cases are shown in Fig. 2. Case 1: If only one emitter is excited by a short coherent $\pi$ pulse at the time marked by a red arrow shown in Fig. 2(a), we can see that the emitter is then in the $\vert e\rangle$ state after the pulse with very high fidelity (about $99\%$ in this example). After that, the excited emitter decays to the ground state through the path $\vert e\rangle \rightarrow \vert f \rangle \rightarrow \vert g \rangle$ and emits a single photon into the waveguide with about $97.7\%$ probability according to the numerical Monte Carlo simulation (Fig. 2(b)).  There are about $2.3\%$ probability that the emitter directly decays back to the ground state and emits a photon into the free space with zero photon in the waveguide. Case 2: If we selectively excite two emitters (e.g., the first and second emitters) by two separate short coherent $\pi$ pulses at the time maked by the red arrow shown in Fig. 2 (c), the system can be in the state $|ee\rangle$ after the pulse with about $97.7\%$ in our numerical example.  The excited emitters then decay back to the ground state through the path $\vert ee \rangle \rightarrow \vert + \rangle \rightarrow \vert ff \rangle \rightarrow \vert gg \rangle$ where $\vert \pm \rangle=\frac{1}{\sqrt{2}}(\vert ef \rangle \pm \vert fe \rangle)$ (Fig. 2(c)) and emit two photons into the waveguide with probability being about $94.2\%$ (Fig. 2(d)). It is clearly seen that the system decays throught the superradiant path since the population in the subradiant state $|-\rangle$ is always zero (black dashed line in Fig. 2(c)). The Monte Carlo simulation shows that the probability of one-photon loss is about $5.6\%$ and two-photon loss is about $0.2\%$ (Fig. 2(d)). Case 3: If three emitters are excited by three short coherent $\pi$ pulses with probability being about $96.5\%$ in our numerical example, they can decay back to the ground state through the path $\vert eee \rangle \rightarrow \vert S_2 \rangle \rightarrow \vert S_1 \rangle \rightarrow \vert fff \rangle \rightarrow \vert ggg \rangle$ where $\vert S_2 \rangle=\frac{1}{\sqrt{3}}(\vert eef \rangle+\vert efe \rangle+\vert fee \rangle)$, $\vert S_1 \rangle=\frac{1}{\sqrt{3}}(\vert eff \rangle+\vert fef \rangle+\vert ffe \rangle)$ as shown in Fig. 2(e). In this process, the system can emit three photons into the waveguide with probability being about $91.2$ (Fig. 2(f)). According to our Monte Carlo simulation, the probability of one-, two- and three-photon losses are about $8.38\%$, $0.38\%$ and 0, respectively. From the photon distribution shown in Fig. 2, we can clearly see that the generated photon states are hihgly nonclassical. 

To clearly see that the photons are indeed emitted by the collective decay, we compare the population decay dynamics of the three cases which is shown in Fig. 3a. We can see that the more emitters are excited the faster the decay rate which is a signature of supperradiant effect. The output photon intensity as a function of time for three different cases are shown in Fig. 3(b) from which we can see that the emitted field intensity is proportional to the number of excited emitters and the photon pulse shape is sharper when more emitters are excited which is another signature of the supperradiant effect. In Fig. 3(c), we show the statistics of the time interval between two successive emitted photons for the case when two emitters are excited. The red dotted lines in Fig. 3(c) and 3(d) are the single photon emission curve for comparison. It is clearly seen that the two emitted photons tend to bunch each other. Similarly, the statistics of the time intervals between the three successive photons for the case when three emitters are excited is shown in Fig. 3(d) from which we can also clearly see the bunching effect among the photons in each bundle.

\begin{figure}[tbp]
\label{Fig4} \centering\includegraphics[width=9cm]{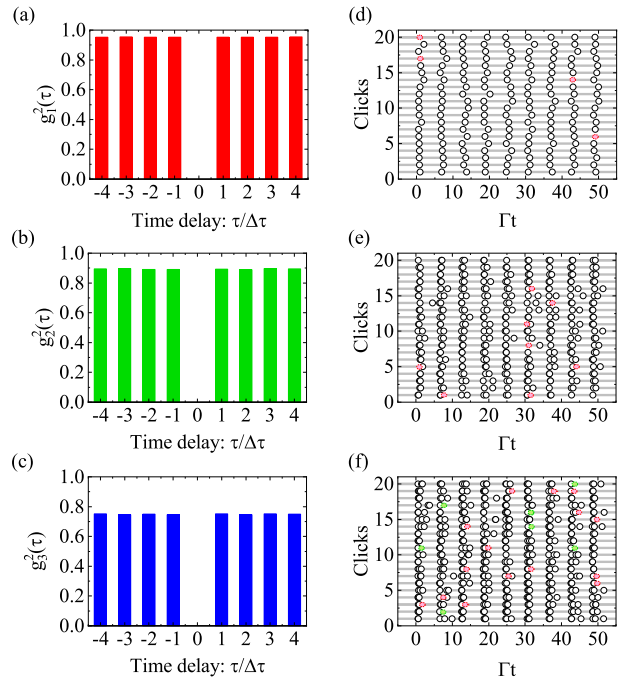}
\caption{(Color online)  Generalized second-order correlation functions driving by 9 pulses (a-c). Clicks over 20 quantum trajectories by quantum Monte Carlo simulation (d-f). (a, d) one-atom excitation, (b, e) two-atom excitation, (c, f) three-atom excitation.}
\label{4}
\end{figure}

To confirm that the photons are indeed emitted by bundles, we calculate the generalized second-order correlation function
\begin{eqnarray}
g_{n}^{2}(\tau)=\frac{\langle a^{\dag n}(0)a^{\dag n}(\tau)a^{n}(\tau)a^{n}(0)\rangle}{\langle (a^{\dag n}a^{n})(0)\rangle\langle (a^{\dag n}a^{n})(\tau)\rangle},
\label{6}
\end{eqnarray}%
which counts the probability that two $n$-photon bundles come successively within the time gap $\tau$. $g_{n}^{2}(0)=0$ indicates that the two $n$-photon bundles never come together. Since we apply pulse excitation instead of continuous excitation, the time gap $0$ here means an interval of time corresponds to a single pulse duration.  The generalized second order correlation functions for the one-, two-, and three-photon cases are shown in Figs. 4(a), 4(b) and 4(c), respectively. From the figures, we can see that $g_{1}^{2}(0)=g_{2}^{2}(0)=g_{3}^{2}(0)=0$. The results clearly show that the photons are indeed emitted by bundles and the bundles are antibunched.

The time structures of the emitted events by 9 consecutive driving pulses with separation $\Delta\tau=6/\Gamma_{1D}$ are shown in Fig. 4(d-f) in which each black circle indicates radiating a single photon and each red circle indicates absence of one photon, and each green circle indicates absence of two photons. The results clearly show the bundle emissions such that one-photon, two-photons, and three-photons event predominate when single, double, and three emitters are excited, respectively. The occasional photon losses are mainly due to the decay channel from $\vert e \rangle \rightarrow \vert g \rangle$ in which the photons are emitted to the free space. 


\section{SUMMARY}

To conclude, we propose a scheme to deterministically generate arbitrary photon number state on demand in an integrated quantum electrodynamics system. Our results show that by selectively pump certain number of emitters into the excited state they can collectively decay to the ground state and emit certain number of photons into the waveguide. The numerical results show that these photons are emitted by bundles and the photons within each bundle is highly correlated. Our study here can find important applications in the on-chip integrated quantum information technology and quantum metrology.


\section*{Acknowledgments}

This work was supported by the National Key R\&D Program of China (Grant No. 2021YFA1400800), the Key-Area Research and Development Program of Guangdong Province (Grant No.2018B030329001), the Guangdong Special Support Program (Grant No.2019JC05X397), the Guangdong Basic and Applied Basic Research Foundation (Grant No. 2023B1515040023), and the Natural Science Foundations of Guangdong (Grant No.2021A1515010039). 

*Email:liaozy7@mail.sysu.edu.cn


\appendix

\section{Master equation}

To prove the validity of our scheme, we can numerically solve the driven master equation \cite{Liao2015,liao2020}.
\begin{eqnarray}
\dot{\rho}_{s}(t)=-i[H_{coh},\rho_{s}(t)]+\mathcal{L}[\rho_{s}(t)],
\label{2}
\end{eqnarray}%
with the coherent Hamiltonian given by
\begin{eqnarray}
H_{coh}(t)=\sqrt{\frac{\gamma}{2}}\sum_{j=1}^{n}(\alpha_{j}(t)\sigma_{eg}^{j}+\alpha_{j}^{\ast}(t)\sigma_{ge}^{j}) \notag \\
+\frac{\Gamma_{1D}}{2}\sum_{jl}sin(k_{a}z_{jl})\sigma_{ef}^{j}\sigma_{fe}^{l},
\label{3}
\end{eqnarray}%
where the first term is the external pumping term and the second term is the coherent atom-atom interaction mediated by the waveguide modes. $\alpha_{j}(t)$ is time-dependent field amplitude of the incident photon pulse. Here, we assume that the pumping photon pulses are the same and they are all short Gaussian pulse which is given by $\alpha_{j}(t)=\sqrt{\frac{\bar{n}_{d}\Delta c}{\pi^{1/4}}}e^{-\frac{\Delta^{2}(x_{0}-ct)^{2}}{2}}e^{ik_{a}x_{0}}$ where $\bar{n}_{d}$ is the average photon number of the driving field, $\Delta$ is the spectrum width, $c$ is the speed of light, and $x_{0}$ is the initial central peak position. Here we assume that the central frequency of the driving field is resonant with the atomic bare frequency. The second term on the right-hand side of Eq. (1) describes the dissipation of the emitter system given by
\begin{eqnarray}
\mathcal{L}[\rho_{s}(t)]&=&-\frac{\Gamma_{1D}}{2}\sum_{j,l=1}^{n}cos(k_{a}z_{jl})\mathcal{L}_{ef}^{jl}[\rho_{s}(t)]\notag \\& &-\frac{\gamma}{2}\sum_{j=1}^{N}\mathcal{L}_{eg}^{jj}[\rho_{s}(t)]-\frac{\gamma_{f}}{2}\sum_{j=1}^{n}\mathcal{L}_{fg}^{jj}[\rho_{s}(t)],
\label{4}
\end{eqnarray}%
where $\mathcal{L}_{ef}^{jl}[\rho_{s}(t)]=\sigma_{ef}^{j}\sigma_{fe}^{l}\rho_{s}(t)+\rho_{s}(t)\sigma_{ef}^{j}\sigma_{fe}^{l}-2\sigma_{fe}^{l}\rho_{s}(t)\sigma_{ef}^{j}$, $\mathcal{L}_{eg}^{jj}[\rho_{s}(t)]=\sigma_{eg}^{j}\sigma_{ge}^{j}\rho_{s}(t)+\rho_{s}(t)\sigma_{eg}^{j}\sigma_{ge}^{j}-2\sigma_{ge}^{j}\rho_{s}(t)\sigma_{eg}^{j}$, and $\mathcal{L}_{fg}^{jj}[\rho_{s}(t)]=\sigma_{fg}^{j}\sigma_{gf}^{j}\rho_{s}(t)+\rho_{s}(t)\sigma_{fg}^{j}\sigma_{gf}^{j}-2\sigma_{gf}^{j}\rho_{s}(t)\sigma_{fg}^{j}$. From the master equation shown in Eq. (2), we can solve the population dynamics of the emitter system.

\section{Quantum Monte Carlo method}

To calculate the photon statistics of the output field, it is convenient to use the quantum Monte Carlo method \cite{zhang2018,plenio1998,molmer1993}. The effective Hamiltonian of the emitter system can be written as
\begin{eqnarray}
H_{eff}(t)=H_{coh}(t)-\frac{i\hbar}{2}\sum_{\beta=R,L,\gamma,\gamma_f}J_{\beta}^{+}J_{\beta}^{-},
\label{5}
\end{eqnarray}%
where $H_{coh}(t)$ is given by Eq. (3) and the effective jumping operators $J_{R,L}^{-}=\sqrt{\frac{\Gamma_{1D}}{2}}\sum_{j=1}^{N}e^{\mp ik_{a}z_{j}}\sigma_{fe}^{j}(t)$, $J_{\gamma}^{-}=\sqrt{\frac{\gamma}{2}}\sum_{j=1}^{N}\sigma_{ge}^{j}(t)$, and $J_{\gamma_{f}}^{-}=\sqrt{\frac{\gamma_{f}}{2}}\sum_{j=1}^{N}\sigma_{gf}^{j}(t)$. Having the effective Hamiltonian, we can use the quantum Monte Carlo method to numerically calculate the dynamics of the system and the photon statistics with the steps as follows:

(1) Suppose the wavefunction at time $t$ is $\vert \varphi(t) \rangle$.

(2) Calculate the probabilities: $P_{R,L}=dt\langle\varphi(t)\vert J_{R,L}^{+}J_{R,L}^{-}\vert \varphi (t)\rangle$ is the probability that the emitters jump from the $\vert e \rangle$ state to the $\vert f \rangle$ state and emit a photon to the right (left) direction; $P_{\gamma}=dt\langle\varphi(t)\vert J_{\gamma}^{+}J_{\gamma}^{-}\vert \varphi (t)\rangle$ is the probability that the emitters jump from the $\vert e \rangle$ state to the $\vert g \rangle$ state and emit a photon to the free space; $P_{\gamma_{f}}=dt\langle\varphi(t)\vert J_{\gamma_{f}}^{+}J_{\gamma_{f}}^{-}\vert \varphi (t)\rangle$ is the probability that the emitters jump from the $\vert f \rangle$ state to the $\vert g \rangle$ state and emit a photon to the free space.

(3) Choose a random number $r$ between 0 and 1, and if
    (a) $r<P_{\gamma}$: $\vert \varphi (t+dt) \rangle=J_{\gamma}^{-}\vert \varphi (t)\rangle/\sqrt{P_{\gamma}/dt}$, a photon emits to the free space;
    (b) $P_{\gamma}<r<P_{\gamma}+P_{R}$, $\vert \varphi (t+dt) \rangle=J_{R}^{-}\vert \varphi (t)\rangle/\sqrt{P_{R}/dt}$ , a photon emits to the right within waveguide;
(c) $P_{\gamma}+P_{R}<r<P_{\gamma}+P_{R}+P_{L}$, $\vert \varphi (t+dt) \rangle=J_{L}^{-}\vert \varphi (t)\rangle/\sqrt{P_{L}/dt}$, a photon emits to the left within waveguide;
(d)$P_{\gamma}+P_{R}+P_{L}<r<P_{\gamma}+P_{R}+P_{L}+P_{\gamma_{f}}$, $\vert \varphi (t+dt) \rangle=J_{\gamma_{f}}^{-}\vert \varphi (t)\rangle/\sqrt{P_{\gamma_{f}}/dt}$, a photon decay to ground states from $\vert f \rangle$.
(e) $r>P_{\gamma}+P_{R}+P_{L}+P_{\gamma_{f}}$, $\vert \varphi (t+dt) \rangle=(I-iH_{eff}dt)\vert \varphi (t)\rangle/\sqrt{1-(P_{R}+P_{L}+P_{\gamma}+P_{\gamma_{f}})}$, no photon is emitted and the system evolves coherently.

(4) Repeat the above procedures for sufficiently large amount of times, we can then obtain the density of the state at arbitrary time by averaging over the results of each trajectory, i.e., $\vert \rho(t) \rangle=\sum_{m}\vert \varphi_{m}(t)\rangle\langle\varphi_{m}(t)\vert$ where $m$ denotes each trajectory and can also calculate the photon statistics including photon counts, photon correlations and so on.


%

\end{document}